\documentclass[aps,pra,english,twocolumn,showpacs,preprintnumbers,amsmath,amshttps://www.overleaf.com/project/65227722c8a1d71888c77233symb,floatfix,superscriptaddress,longbibliography]{revtex4-1}

\usepackage[T1]{fontenc}
\usepackage{graphicx}
\usepackage{amssymb}
\usepackage{babel}
\makeatletter

\usepackage[version=3]{mhchem}

\usepackage{color}
\newcommand{\revisedtext}[1]{\textcolor{black}{#1}}

\makeatother
\usepackage{babel}
\makeatother

\begin{document}

\author{M.  Bostr{\"o}m}
\email{mathias.bostrom@ensemble3.eu}
\affiliation{Centre of Excellence ENSEMBLE3 Sp. z o. o., Wolczynska Str. 133, 01-919, Warsaw, Poland}
\affiliation{Chemical and Biological Systems Simulation Lab, Centre of New Technologies, University of Warsaw, Banacha 2C, 02-097 Warsaw, Poland}

\author{A. Gholamhosseinian}
%\email{ayda.gholamhosseinian@mail.um.ac.ir}
\affiliation{Department of Physics, Ferdowsi University of Mashhad, Mashad, Iran}

\author{S. Pal}
\affiliation{Centre of Excellence ENSEMBLE3 Sp. z o. o., Wolczynska Str. 133, 01-919, Warsaw, Poland}

\author{Y. Li}
\affiliation{Department of Physics, Nanchang University, Nanchang 330031, China}
\affiliation{Institute of Space Science and Technology, Nanchang University, Nanchang 330031, China}

\author{I. Brevik}
\email{iver.h.brevik@ntnu.no}
\affiliation{Department of Energy and Process Engineering, Norwegian University of Science and Technology, NO-7491 Trondheim, Norway}

\title{Semi-Classical Electrodynamics and the Casimir Effect}
\date{\today}

\begin{abstract}
From the late 1960s and onwards the groups of Barry Ninham and Adrian Parsegian, and their many collaborators, made a number of important contributions to theory and experiment of intermolecular forces. In particular, they explored the semi-classical theory: Maxwell’s equations and Planck quantization of light $\rightarrow$ Lifshitz and Casimir interactions. We discuss some selected thought-provoking results from Ninham and his group. Some of the results have been conceived as controversial but, we dare, say never uninteresting.
\end{abstract}
%\abstract{}
%\date{\today}

\maketitle

\section{Introduction}
\label{Intro}
\par Since the prediction of the Casimir effect in 1948 and its experimental confirmation in the period after that, there has been a significant interest in studying the forces caused by fluctuations both theoretically and experimentally~\cite{Casi,CasimirPolder48}. Before reviewing some contributions to Casimir physics from semi-classical electrodynamics theory, \revisedtext{that is Maxwell’s equations and Planck quantization of light leading to Lifshitz and Casimir interactions}, (with particular emphasis on the work by Barry Ninham and co-workers), we first present some historical reflections. \revisedtext{The history of intermolecular forces actually goes back to the early history of science.}
Thomas Young notably wrote an article on molecular forces in 1805~\cite{Young,Craik}. Young deduced that they had to obey $1/r^6$-power-laws. Rev. Challis of Trinity College in a major address to the British Association \revisedtext{1836\,\cite{Challis}}, reviewed the state of molecular forces between colloidal particles, suggested interferometry for direct measurements, quoting Fresnel, and referred to the subject as "this the highest Department of Science" for which he coined the term Mathematical Physics. The famous article by J. Clerk Maxwell in the 9th edition of Encyclopaedia Britannica (1876) discussed capillary action and molecular forces, updated by Lord Rayleigh in the 11th edition (1911).
\revisedtext{Ruder Boscovic\,\cite{boskovic1966theory},} a Croatian Jesuit Priest based in Rome after whom the Ruder Boscovic Institute in Zagreb is named, developed a system of the world essentially inventing statistical mechanics. His book appeared around 1600. To do so, he needed a molecular potential. His effective potential oscillated with the period a molecular diameter tailing off into a gravitational $1/r^2$. The basis of the study of intermolecular forces was laid by Johannes Diderik van der Waals in 1873~\cite{van1873over}. He clarified the concept of interparticle forces and how molecules interact. Quantum fluctuations create intermolecular forces that exist throughout macroscopic bodies. At the molecular separations of about a few nanometres or less, these interactions causing the attraction and repulsion between molecules, are the familiar van der Waals forces.

\par As discussed by Derjaguin, Abrikossova, and Lifshitz in their 1956 review paper~\cite{Derja}, the correct understanding of the nature of molecular forces was initially proposed by P. N. Lebedew, back in 1894~\cite{Derja,Lebedev2}: "There exist intermolecular forces whose origin is closely connected with radiation processes." In general, it is necessary to understand van der Waals forces and understand their importance compared to other molecular forces.

\par Fundamental and applied research on molecular forces continued until Fritz London proposed the general theory of molecular forces in 1930~\cite{london1930theorie}. This theoretically improved the understanding of molecular dispersion forces, and contributed to the interpretation of van der Waals forces and other molecular forces. Also, the significant contributions of Evgeny Lifshitz should be addressed. In 1955, Evgeny Lifshitz explained how the oscillating charge distribution in molecules leads to the creation of attractive forces~\cite{lifshitz1955theory}. These explanations contribute to a deeper understanding of molecular forces and their role in various phenomena.

\par \revisedtext{Ever since the Russian researchers Derjaguin and Abrikossova~\cite{Derja} performed their force measurements, there has been a strong focus on the phenomena The first set of experiments, notably measuring interactions between quartz and metal plates, studied only the so-called retarded region}. Experiments of Tabor and Winterton~\cite{Tab} and subsequently of Israelachvili and Tabor~\cite{IsraTabor,White} fitted the \revisedtext{measured force} to a power law \revisedtext{function of $1/L^{p}$ (where $L$ is the distance), where $p$ varied from non retarded ($p = 3$) to fully retarded ($p = 4$) value. In these early experiments, a gradual transition was observed from non-retarded  to retarded interaction, as the distance between the surfaces increased from around 12 nm upto 130 nm~\cite{IsraTabor}}. Surface force measurements~\cite{Derja,TabNature,Tab,IsraTabor,White} and theoretical clarifications and extensions of the Lifshitz formula~\cite{Dzya,ParNin1969,NinhamParsegianWeiss1970,NinPars1970,Maha} to include, for example, magnetic~\cite{Richmond_1971} and conducting particles~\cite{RichDavies,Davies}, and liquids between unequal surfaces,  were pioneered in the 1970s by the group of Barry Ninham, Jacob Israelachvili, and their co-workers at the Department of Applied Mathematics (Australian National University).

\par According to the \revisedtext{fundamental theory}, the Lifshitz force can also be repulsive, which is an interesting feature that has attracted a lot of attention~\cite{Dzya,Rich1,Rich}. Anderson and Sabiski demonstrated this phenomenon in their research on liquid helium films on smooth surfaces of calcium fluorite ($\text{CaF}_2$), among other similar molecularly smooth surfaces~\cite{AndSab}. The thickness of the films in the experiment ranged from 10-200 \AA, and could be measured with an accuracy of a few percent in most cases.  Several past publications by Barry Ninham have explored the history of intermolecular forces in more detail. The book by Barry W. Ninham and Pierandrea Lo Nostro is particularly interesting~\cite{Ninhb}.

\par We focus first on reviewing a work that our close and distinguished collaborator Prof. Barry W. Ninham wrote in 1970 together with Adrian V. Parsegian and G. H. Weiss~\cite{NinhamParsegianWeiss1970}. The reason to highlight this article is that we feel it is not well enough recognized in the field. The theories of intermolecular dispersion forces have occupied such a vast literature that one would suspect very little should remain to be said. However, even lately, new applications of the fundamental theory have arisen. We will first address the semi-classical theory itself and then briefly discuss our contribution to the theory of Casimir interaction between real metal plates at high temperatures/large separations. \revisedtext{A controversial, as well as very intriguing, idea is briefly explored in the current work highlighting that the high-temperature Casimir effect might have a role even in nuclear physics~\cite{Ninham_Brevik_Bostrom_2022}. To be more specific: it was shown in an unpublished note by Ninham and Pask more than 50 years ago how Maxwell's equations for the electromagnetic field with Planck quantization of allowed modes appear to provide a semiclassical account of nuclear interactions. The direct consequence if this idea has any relevance is that mesons would emerge as plasmons, collective excitations in an electron-positron pair sea~\cite{Ninham_Brevik_Bostrom_2022}.} We then proceed to present some work that was initiated by Ninham around 1970 on excited state interaction between atoms~\cite{Bostrom1}.  \revisedtext{Related self energies and excited state interactions are, for example, important in photobiochemistry.} We finally wrap up our story with a few concluding words.

\section{The Ninham, Parsegian \& Weiss semi-classical derivation of Lifshitz theory}

The theory due to Lifshitz was readdressed by Dzyaloshinskii et al.~\cite{Dzya} via some lengthy arguments that exploited Green's function techniques in quantum field theory. We will outline the general ideas behind the much simpler semi-classical theory of dispersion interactions. The paper by Ninham et al.~\cite{NinhamParsegianWeiss1970}, \revisedtext{which} we will follow in this section with some changes in notation, expanded on general ideas presented by van Kampen et al.~\cite{VANKAMPEN1968307}. \revisedtext{In this section we point out that van Kampen et al. only considered the zero-temperature and non-retarded limit. We use the electrodynamics boundary conditions given by Jackson~\cite{Jackson:100964}} that the components of (${\bf{E}}_\omega,{\bf{H}}_\omega$), $E_x$, $E_y$, $\varepsilon E_z$, $H_x$, $H_y$, and $\mu H_z$ are continuous \revisedtext{at interfaces between the media, parallel} to the $xy$-plane at $z = 0$ and $z = d$. \revisedtext{We are thus considering the simple case of two half-spaces interacting across a media.}
We assume that the dielectric \revisedtext{permittivities}  are different, $\varepsilon_1(\omega)\neq\varepsilon_2(\omega)$, and the magnetic \revisedtext{permeabilities}  equal to one. The solutions  (${\bf{Y}}=\sum_{\omega} {\bf{Y}}_\omega e^{-i\omega t}$ where ${\bf{Y}}_\omega$ $\in \{$ ${\bf{H}}_\omega$,  ${\bf{E}}_\omega\}$) have \revisedtext{normal mode frequencies from} the wave equations~\cite{NinhamParsegianWeiss1970},
\begin{equation}
\nabla^2{\bf{Y}}_\omega+(\varepsilon \omega^2/c^2){\bf{Y}}_\omega=0,
\end{equation}
together with $\nabla\cdot{\bf{Y}}_\omega=0$.  For the separate components of ${\bf{E}}_\omega$ and ${\bf{H}}_\omega$, one assumes the form~\cite{NinhamParsegianWeiss1970} $\Theta(z) e^{i (ux+vy)}$, where $u$ and $v$ are the wave vector components parallel to the surface, and $\Theta^{\prime\prime}(z)=\gamma^2 \Theta(z)$.
\revisedtext{Here $\gamma^2=\kappa^2-[\omega^2 \varepsilon(\omega)/c^2]$ and $\kappa=\sqrt{u^2+v^2}$ is the real component of the wave vector parallel to the slab of the intermediate film.  Ninham, Parsegian and Weiss showed that normal modes $(\omega_i)$ are solutions of transverse magnetic (TM)  and transverse electric (TE) dispersion relations~\cite{NinhamParsegianWeiss1970},}
\begin{equation}
  D_1(\omega;d)=1-\Delta^2_{\rm TM} e^{-2 \gamma_2 d}=0= D_2(\omega;d)=1-\Delta^2_{\rm TE} e^{-2 \gamma_2 d},
  \label{NPWeq6}
\end{equation}
where~\cite{NinhamParsegianWeiss1970}
\begin{equation}
    \Delta_{\rm TE}= \frac{\gamma_2-\gamma_1}{\gamma_2+\gamma_1};  \,\,\,\,\, \Delta_{\rm TM} = \frac{\varepsilon_1\gamma_2-\varepsilon_2 \gamma_1}{\varepsilon_1\gamma_2+\varepsilon_2 \gamma_1}.   \label{NPWeq7}
\end{equation}
%The modes are of two types, TE and TM and there are no TE evanescent modes.
\revisedtext{The two types of electromagnetic modes are TE and TM, and there are no TE evanescent modes.} This is put into question for the Drude-plasma model for real metal surfaces as discussed briefly in the next section.
\revisedtext{The requirement of the surface-type solutions (those that are well-behaved and vanish at infinity), Re($\gamma_2)>0$ implies that $\kappa\geq(\omega/c){\rm Re}(\varepsilon_2^{1/2})$ in what follows. Ultimately the fundamental
dispersion relations in Eq. (\ref{NPWeq6}) combine into one simple relation: $D=D_1 D_2 =0$~\cite{NinhamParsegianWeiss1970}}.
\revisedtext{The Gibbs interaction free energy is given by\cite{NinhamParsegianWeiss1970},}
\begin{equation}
F(d,T) = \frac{1}{2 \pi} \int_0^\infty [F_d(r)-F_\infty(r)] r dr,
\label{NPWeq9}
\end{equation}
the next step is to integrate over wave vector ($\mathrm{r}$),
\begin{equation}
    F_d(r) =  k T \sum_{j}\ln[2    \sinh(\beta \hbar \omega_j(r)/2)],
\end{equation}
here in the subsequent steps, the sum must be taken over all the available real roots of Eq.~\eqref{NPWeq6}. From this it follows\,\cite{NinhamParsegianWeiss1970}
\begin{equation}
\sum_{j} g(\omega_j)=\frac{1}{2 \pi i} \oint_C g(\omega) [1/D(\omega)] [dD(\omega)/d\omega]d\omega.
\label{NPWeq11}
\end{equation}
\revisedtext{In the formula, the path $C$ includes the important points where $D$ has zeros but doesn't include the poles of this function (for details see the work by Ninham, Parsegian, and Weiss\,\cite{NinhamParsegianWeiss1970}). To ensure the validity of Eq. (\ref{NPWeq11}), the functions $g(z)$ and $D(z)$ must exhibit analyticity or smoothness,}
where the contour $C$ includes the relevant zeros of $D$, and excludes poles of $g(\omega)$ (\revisedtext{for details see the work by Ninham, Parsegian and Weiss\,\cite{NinhamParsegianWeiss1970}}).
Note that $g(z)$ and $D(z)$ are assumed analytic for Eq. (\ref{NPWeq11}) to hold.
Since \revisedtext{$g(\omega)=\ln[2 \sinh(\beta \hbar \omega/2)]$ has branch cuts}, it is convenient to expand it as\,\cite{NinhamParsegianWeiss1970}
\revisedtext{\begin{equation}
\begin{aligned}
g(\omega)&=\ln[2 \sinh(\beta \hbar \omega/2)] \\
&= \ln\Big[ e^{\frac{\beta \hbar \omega}{2}}- e^{-\frac{\beta \hbar \omega}{2}} \Big]\\
&= \frac{\beta \hbar \omega}{2} + \ln\Big[ 1- e^{-\beta \hbar \omega}\Big]\\
&= \frac{\beta \hbar \omega}{2}-\sum_{n=1}^{\infty} \frac{1}{n} e^{-n \beta \hbar \omega}
\label{NPWeq12}
\end{aligned}
\end{equation}}
and consider each term separately. \revisedtext{To proceed formally, we choose the path in Eq.~\eqref{NPWeq11} starting from $-i\infty$ to $i\infty$ along the imaginary axis, then move around the right half-plane along a semicircular path with an infinity radius.} Since $\varepsilon(|\omega|)\rightarrow 1$ as $|\omega|\rightarrow \infty$, $D(|\omega|)=1$ on
the semicircle and one can write~\cite{NinhamParsegianWeiss1970}

\revisedtext{\begin{eqnarray}
F_d(r)=\frac{1}{2\pi i} \int_{\infty}^{-\infty} g(i\xi) \frac{d\ln D(i\xi;d)}{d\xi}d\xi\\
=\frac{\hbar}{2} \sum_{j} \omega_j+\frac{\hbar}{2 \pi} \sum_{n=1}^{\infty} \int_{-\infty}^{\infty}\cos(n \beta \hbar \xi)\ln D_r(i\xi;d)d\xi \nonumber\\
-\frac{i \hbar}{2 \pi} \sum_{n=1}^{\infty} \int_{0}^{\infty}\sin(n \beta \hbar \xi)\ln\bigg[\frac{D_r(i\xi;d)}{D_r(-i\xi;d)}\bigg]d\xi.
\label{NPWeq14}
\end{eqnarray}}
Using standard mathematics we exploit the identity~\cite{NinhamParsegianWeiss1970}
\begin{equation}
\sum_{n=1}^{\infty}\cos(n x)=\pi\sum_{n=-\infty}^{\infty} \delta(x-2\pi n)-\frac{1}{2}.
\label{NPWeq15}
\end{equation}
When the delta functions are substituted into the integrals, the integrations can be carried out ~\cite{NinhamParsegianWeiss1970}
\begin{eqnarray}
\frac{\hbar}{2} \sum_{j} \omega_j+\frac{\hbar}{2 \pi} \sum_{n=1}^{\infty} \int_{-\infty}^{\infty}\cos(n \beta \hbar \xi)\ln D_r(i\xi;d)d\xi \\
= \frac{kT}{2}\sum_{n=-\infty}^{\infty}\ln D_r(i\xi_n;d),
\label{NPWeq16}
\end{eqnarray}
where the Matsubara frequency $\xi_n=2 \pi k T n/\hbar$ so that~\cite{NinhamParsegianWeiss1970}
\begin{eqnarray}
F_d(r)
=
\frac{kT}{2}\sum_{n=-\infty}^{\infty}\ln D_r(i\xi_n;d)
 \\
- \frac{i \hbar}{2 \pi} \sum_{n=1}^{\infty} \int_{0}^{\infty}\sin(n \beta \hbar \xi)\ln\bigg[\frac{D_r(i\xi;d)}{D_r(-i\xi;d)}\bigg]d\xi.
\label{NPWeq17}
\end{eqnarray}
Noteworthy for dielectric functions depending on $\omega^2$ only, one term in Eq.~\eqref{NPWeq17} turns out to be zero by symmetry and Eq.~\eqref{NPWeq17} reduces to~\cite{NinhamParsegianWeiss1970}
\begin{equation}
F_d(r) = kT \sum_{n=0}^{\infty}{}' \ln D_r(i\xi_n;d)
\label{NPWeq18}
\end{equation}
\revisedtext{The prime notation represents that the term corresponding to n = 0 is multiplied by 1/2.}
where the prime indicates that the n = 0 term is multiplied by 1/2. This equation is, for instance, valid for the so-called plasma model discussed in the next section.

\section{A high-temperature semi-classical application: the Drude-Plasma controversy}
\par The Casimir interaction between real metal surfaces has caused controversy in the field of Casimir physics. The 1970 paper by Ninham and co-workers discussed above~\cite{NinhamParsegianWeiss1970}  could be relevant to this problem. As is well known a real metal has a finite static conductivity~\cite{Bost2000,Ser2018}. The so-called Drude model is a suitable model for the optical and dielectric properties of a real metal for small frequencies. The dielectric function can be described within the Drude model as
\begin{equation}
\varepsilon \left( \omega  \right) = 1 + \frac{{4\pi i\sigma \left( \omega  \right)}}{\omega } = 1 - \frac{{\omega _{pl}^2}}{{\omega \left( {\omega  + i\gamma } \right)}}.
\label{DrudeEq}
\end{equation}
Putting the dissipation parameter $\gamma $, equal to zero is in the literature called using the plasma model,
\begin{equation}
\varepsilon \left( \omega  \right) = 1 - \frac{{\omega _{pl}^2}}{{{\omega ^2}}}.
\label{plasmaEq}
\end{equation}
Notably, for metal surfaces using the Drude dielectric function, the condition used to pass from Eq.~\eqref{NPWeq17} to Eq.~\eqref{NPWeq18} is not fulfilled. However, the dissipation parameter has an actual physical basis. It is not zero for any real metal. \revisedtext{Its origin is via scattering of carriers against impurities within the lattice. Importantly, when} using the plasma model, one by simple \revisedtext{ad hoc decision} neglects these effects and \revisedtext{as a result} the static conductivity becomes infinite. Bostr{\"o}m and Sernelius~\cite{Bost2000}  \revisedtext{demonstrated how including} the dissipation parameter has a \revisedtext{critical} effect on the \revisedtext{predicted} interaction at \revisedtext{large surface} separations, where  \revisedtext{temperature effects impact the results strongly}. The plasma model predicts a result coinciding with that of the classical Gedanken experiment by Casimir between two perfectly reflecting half-spaces, while the Drude model predicts that this result is reduced by a factor of two. To understand how these drastic effects occur, we need to look at the TM and TE normal modes involved in the problem. When dissipation is included, there are also TE evanescent modes. At separations where the temperature is important, the contribution to the interaction from these TE evanescent modes completely cancels those from the TE propagating modes.
\revisedtext{It is known that experimental results in general agree better with the zero-temperature results between real metal surfaces and very well with the zero-temperature result for planar ideal metal surfaces (i.e. the so-called Casimir Gedanken experiment).  The agreement is better still with the theoretical room-temperature result obtained when using the so-called plasma model. This was the seed of the longstanding controversy in the field.  The finite temperature Casimir effect between metallic surfaces is a complex phenomenon and care has to be taken about the electrostatic patch potentials, which have caused uncertainties in the actual interpretation of the data in experiments. Different theoretical groups have found fundamentally different results\,\cite{Bost2000,Bord}. A particularly useful aspect of the original Lamoreaux experiment \cite{Lamo1997} was that it was carried out at large enough separations where finite temperature corrections can be expected. Sushkov and Lamoreaux with collaborators later presented results using a similar experimental setup where separations were varied from 0.7 to 7 $\mu$m\,\cite{SushNP}. The theoretical predictions based upon the Drude model were found to agree with the observed results to
acceptable accuracy. It is important to point out, however, that other experiments\,\cite{HarrisPhysRevA.62.052109,DeccaPhysRevLett.91.050402,RevModPhys.81.1827,Mohid,KlimchitskayaMohideenMostepanenko} (more references in the recent work by Vladimir Mostepanenko and Galina Klimchitskaya in this special issue\,\cite{MostKlim2023}), yielded results in good accordance with the plasma model rather than the Drude model. The reason for contradictory results (both theoretical and experimental) is not known to the authors of the present review.  There is still a need for more experiments and theoretical analysis focusing on Casimir-Lifshitz forces in different systems that include \revisedtext{interacting conducting (metallic) objects}. However,} it is not the purpose of the current work to explore this problem. For one side of this long story (and relevant references), we refer the readers to a very recent article by Mostepanenko and Klimchitskaya~\cite{Bordagbook,MostKlim2023}. For another side of the story, one could for example consult the very clear account by Sernelius~\cite{Ser2018}. Additional information can be found in the literature\,\cite{milton01,Ser,Buhmann12a,Buhmann12b}. But, perhaps, a correct calculation for high-temperature/large-separation Casimir force between real metal surfaces has still not been carried out. The solution might {\revisedtext{from a fundamental point of view, if perhaps not necessary from a practical point of view,}} involve both the use of our Eq.~\ref{NPWeq17} and the inclusion in the theory of any intervening plasma as in the next section.

\section{Another intriguing semi-classical story:  Casimir interaction energy across a plasma}
\label{NinhamParsegianWeiss}
\par
\revisedtext{Researchers have been looking into Casimir forces over time because of their fundamental role in electron stability, particle physics, and nuclear interactions~\cite{PhysRevA.67.030701,CASIMIR1953846,BoyerPhysRev.174.1764}. We recently looked at a Casimir-Yukawa problem that's similar to the classic story of electron stability, often known as "the Casimir mousetrap"~\cite{BoyerPhysRev.174.1764}. This problem explores how negative charges on an electron surface create a repulsive force between surface parts, which has to be counteracted by an attractive force to retain a finite electron radius. Casimir proposed that the attractive Poincar{\'e} stresses could be caused by the zero-point energy present in electromagnetic vacuum fluctuations~\cite{BoyerPhysRev.174.1764}. Nevertheless, the work of Boyer and others showed that although the interaction's magnitude was correct, it had the wrong sign and resulted in a repulsive force~\cite{BoyerPhysRev.174.1764}.} \revisedtext{Other relevant models, such as the dielectric ball, also exhibit their respective problems, some of which are still triggering discussions recently.} Around 50 years ago, Ninham and Pask (unpublished) found that the zero-temperature Casimir vacuum fluctuation energy was enough to provide the binding energy of nucleons in a nucleus. At finite temperatures, the expression discussed in previous sections became~\cite{Dzya},
\begin{equation}
F(d,T) = \frac{kT}{\pi} \sum_{n=0}^\infty{}' \int_0^\infty dq \, q \ln[1 - e^{-2d\sqrt{q^2 + \xi_n^2/c^2}}].
\end{equation}
Explicitly, \revisedtext{in vacuum (i.e. in the complete absence of an intervening electron-positron  plasma)}, Ninham and Daicic derived the following useful expansion~\cite{,PhysRevA.57.1870,PhysRevA.67.030701,EPJDNinham2014,Ninham_Brevik_Bostrom_2022},
\begin{equation}
F(d,T)\approx \frac{- \pi^2\ \hbar c}{720d^3}- \frac{\zeta(3) k^3 T^3}{2 \pi\hbar^2 c^2}+\frac{\pi^2 d k^4 T^4}{45\hbar^3 c^3} +\cdots,		
\end{equation}
where $\zeta(3)\approx1.202$. We observe that the initial term corresponds to the attractive zero temperature Casimir result. The third term in this expression corresponds to a black body radiation energy (in vacuum and at equilibrium). Ninham and Bostr\"om discussed more than 20 years ago how this term opposes the attractive Casimir term~\cite{PhysRevA.67.030701}.
%Following our previous work we assume that the first and third terms are equal at equilibrium leading to a very high temperature.
The remaining term is a chemical potential term that in the Gibbs free energy is well recognized as being due to an electron-positron plasma formed from the photons \revisedtext{inside the nuclear gap} $(e^-+e^+\leftrightarrow\gamma$)~\cite{LandauLifshitzStatPhys1}. The second term can be analyzed using the known density of an electron-positron plasma~\cite{LandauLifshitzStatPhys1},
\begin{equation}
\rho_-+\rho_+=\frac{3 \zeta(3) k^3 T^3}{ \pi^2\hbar^3 c^3}.
\end{equation}
For a pair of perfectly conducting plates, the Casimir interaction energy across \revisedtext{an electron-positron plasma is},
\begin{equation}
F(d,T) = \frac{{kT}}{\pi }\sum\limits_{n = 0}^\infty  {'\int_0^\infty
d } qq\ln \left[ {1 - {e^{ - 2d\sqrt {{q^2} + {{({\xi _n}/c)}^2} +
{\kappa ^2}}
}}} \right],\label{EqA11}
\end{equation}
recall that $\kappa=\omega_p/c$. For any separation at high enough temperatures, or for any finite temperature at large enough separations, it follows~\cite{PhysRevA.57.1870,PhysRevA.67.030701,EPJDNinham2014} an expansion of the form,
\begin{equation}
F(d,T)=-\frac{kT\kappa}{4\pi} \frac{e^{-2\kappa d}}{d} \Big[1+\frac{1}{2d\kappa}\Big]-\frac{(kT)^2 e^{-2 \eta d}}{\hbar c}\frac{e^{-\rho^*\eta d}}{d}+ O(e^{-4 \eta d}),
\end{equation}
where $\rho^\ast=\rho\ e^2 \hbar^2/(\pi m_e k^2\ T^2)$, $\eta=2kT/(\hbar c)$ and $\kappa$ is defined above. Both the $n=0$ and $n>0$ terms behave similarly to the Yukawa potential~\cite{EPJDNinham2014,Ninham_Brevik_Bostrom_2022}. Both provide contributions to the Casimir-Yukawa binding energy, depending on the separation ($\sim$0.9\,MeV from the n=0 term and $\sim$3.6\,MeV from the n>0 term), and surprisingly close to the experimentally observed binding energy per nucleon (1.1\,MeV $\sim$ 8.8\,MeV)~\cite{Ninham_Brevik_Bostrom_2022}. \revisedtext{The idea that there ought to be some kind of link between electromagnetic and nuclear forces goes back to Richard Feynman (via private communication in the late 1960's between Barry W. Ninham and Freeman Dyson who told Ninham that Feynman had believed there ought to be a connection between electromagnetic theory and nuclear interactions). This idea was first explored in a yet unpublished work by Ninham and Pask from early 1970s. We have revived and expanded on this idea in a series of publications\,\cite{PhysRevA.57.1870,PhysRevA.67.030701,EPJDNinham2014}. The explicit derivation of meson mass, nuclear binding energy and lifetimes\,\cite{NinhamPhysRev.145.209} were recently discussed at length by Ninham {\it et al.}\,\cite{Ninham_Brevik_Bostrom_2022}.}

\section{Semi-classical derivation of resonance interaction between exited state atom pair}
\par The semi-classical formalism was also able to describe in detail the ground state van der Waals potentials between a pair of molecules, or between a molecule and a surfaces~\cite{Maha}. Here, as one slightly more controversial example~\cite{McLachlan,Avery,Power2,Bostrom1,Andrews}, we explore what predictions come out from semi-classical theory for the resonance interaction energy between two identical atoms in an excited configuration. The results in this section were, in the zero-temperature limit, derived about 50 years ago by Ninham, John Mitchell, and others at the Department of Applied Mathematics (Australian National University), and finally, after deep contemplation and a final extension to finite temperatures~\cite{Bostrom1}, published 20 years ago. Notably, the results are, in the non-retarded limit, identical to the perturbation theory results~\cite{Avery,Power2,Andrews,AndrewsCurutchetScholes2011}, but in the retarded and finite-temperature limits, non-oscillatory results are found. This contrast against the oscillatory long-range retarded resonance interaction obtained from perturbation theory~\cite{Avery,Power2}.

 The normal mode expression used to calculate ground state van der Waals interactions in the case of two identical atoms in air, \revisedtext{\begin{equation}
1-\alpha(1|\omega) \alpha(2|\omega) T(d|\omega)^2=0,
\label{Eq3}
\end{equation}
can be separated into one anti-symmetric and one symmetric part.} \revisedtext{Here $T(\rho|\omega)$ is the field susceptibility~\cite{McLachlan} in a material with dielectric function $\varepsilon(\omega)$, and $\alpha(j|\omega)$ represented the polarizability of atom $j$.}  The excited symmetric state has a substantially shorter lifetime than the excited anti-symmetric state, which can cause the system to end up in an excited anti-symmetric state~\cite{Bostrom1}. The first-order dispersion energy of such an anti-symmetric state comes from~\cite{Bostrom1},
\begin{equation}
\label{Eq5}
U(d)= \hbar [\omega_{r} (d)-\omega_{r} (\infty)].
\end{equation}
The solution of Eq.~\eqref{Eq3} is the pole of the anti-symmetric part (of the underlying Green's function). \revisedtext{We changed the integration path around this pole to obtain an expression} for the first order excited state resonance interaction energy~\cite{Bostrom1},
\begin{eqnarray}
U(d) = (\hbar/ \pi) \int_0^\infty d \xi \ln[1+\alpha(1|i \xi) T(d|i \xi)].
\label{Eq6}
\end{eqnarray}
As pointed out in the past, any finite temperature systems can approximately be dealt with in the same way as for the corresponding ground state problem~\cite{PhysRevA.57.1870,Daicic2}. The temperature ($T$) dependence follows when replacing the integration over imaginary frequencies by a summation over discrete frequencies~\cite{lifshitz1955theory}. The leading term, at large separation when the modes in the ($\pm$;$x$) branch are excited, is~\cite{Bostrom1}
\begin{equation}
U(d,T)\simeq \pm {{2 k_B T} \over { d^{3}}} \sum_{n=0}^{\infty}{'} \alpha(i \xi_n) e^{-x n} [1+x n+x^2 n^2],
\label{Eq35}
\end{equation}
where $x=2 \pi k_B T d/(\hbar c)$. In a standard way we approximate the polarizability with $\alpha (0)$ at large enough separations. Within this approximation, the resonance free-energy is~\cite{Bostrom1},
\begin{equation}
\begin{array}{l}
U(d ,T) \simeq  \pm \frac{{2{k_B}T\alpha (0)}}{{2{d ^3}{{({e^x} - 1)}^3}}}
[1 + {e^{3x}} - {e^x}(1 + 2x - 2{x^2})  \\+ {e^{2x}}( - 1 + 2x + 2{x^2})].
\end{array}
\label{Eq35b}
\end{equation}
For small values of $x$, this free energy of resonance interaction goes as $1/d^4$. However, for any finite temperature, the long-range interaction within the Ninham model is dominated by the $n=0$ term. This term is here~\cite{Bostrom1}
\begin{equation}
U(d ,T)_{n=0}=\pm k_B T \alpha(0)/d^3.
\label{Eq36}
\end{equation}
 This manifestation of the correspondence principle is identical in nature to the result obtained for the retarded van der Waals interaction between two ground-state atoms~\cite{PhysRevA.57.1870,Daicic2}. \revisedtext{This highlights that the quantum nature of light has an important role behind the softening of intermolecular interactions between ground state or excited state atoms (and indeed in the same way for Lifshitz interactions between macroscopic surfaces).}

\section{Discussions and Future Outlooks}
\par This concise review \revisedtext{primarily} aims to engage an insightful discussion concerning the semi-classical theory of interactions with ground and excited state van der Waals, Lifshitz, and Casimir forces. This work \revisedtext{also aims to shed} new light on a small, but missing, element of information that might provide some understanding to settle the Drude-plasma controversy. To be more precise, as we have already discussed in the work by Ninham, Parsegian, and Weiss, the conventional Lifshitz theory left out one extra term. Further investigations of the Casimir effect at high-temperatures/short-range regimes may offer more evidence for its potential influence in \revisedtext{both} meson and atom-atom physics.
\revisedtext{The modern research on \revisedtext{van der Waals,} Lifshitz, and Casimir interactions was pioneered by Ninham and Parsegian more than 50 years ago. Their respective groups demonstrated how to use the complicated Lifshitz theory and how to derive it in a much-simplified way. Researchers from around the world have developed the field for the last 50-60 years \revisedtext{(notably Russian researchers, including Barash and Ginzburg\,\cite{BARASH1989389}, presented some classic works that are similar to those that came from Ninham's group in Australia)}. However, it is still a very active research field\,\cite{Bost2000,MostKlim2023}.
The field was for instance energized by Elbaum and Schick who predicted that ice can have a nanometer sized premelted water layer on an ice surface caused by van der Waals, Lifshitz, intermolecular forces\,\cite{Elbaum}. In general surface charges, ions, and impurities can induce water films many orders of magnitude thicker\,\cite{Elbaum1993,Wilen,Wettlaufer}. The effects caused by ionic interactions are in general very complicated, due partly to the fact that the polarizability of ions leads to a non-linear coupling of van der Waals and ionic forces leading to the macroscopic double-layer and Lifshitz forces acting across salt solutions\,\cite{ParsonsNinham2009,Pars,ParsonsNinham2012,DuignanNinhamParsons2013solvation,ParsonsSalis2015,ThiyamFiedlerBuhmannPerssonBrevikBostromParsons2018}.
An impact from such intermolecular forces has also been proposed for frost heaving\,\cite{WilenFrost} and thunder cloud charging\,\cite{Dash1995,DW,doi:10.1098/rsta.2018.0261}. Ice melting at surfaces and interfaces could be relevant as habitats for life on planets and moons in permafrost regions but also on other planets and moons in the solar system and beyond\,\cite{Boxe} (for useful discussions on planetary science see references\,\cite{dePater,Catling,EnceladusBook}).  The reverse reaction with ice forming on a water surface via Lifshitz interactions was ruled out by Elbaum and Schick\,\cite{Elbaum2}.  However, in contrast, recent re-investigations of the optical properties of water and ice suggest such a role\,\cite{JohannesWater2019,LuengoMarquez_IzquierdoRuiz_MacDowell2022}. Following along these lines we investigated how the Lifshitz interaction can contribute to some geophysical effects including ice layer formation on gas hydrate surfaces\,\cite{BostromEstesoFiedlerBrevikBuhmannPerssonCarreteroParsonsCorkery2021}. We have recently proposed such dispersion interactions as potential energy sources behind a secondary ice growth mechanism on partially melted ice clusters within mist, fog, and potentially also in clouds\,\cite{BostromvdWicegrowthmistPCCP2023}. The contributions from intermolecular forces to geophysics is an evolving research field with important contributions from McDowell and co-workers\,\cite{LUENGOMARQUEZMacDowell2021,LuengoMarquez_IzquierdoRuiz_MacDowell2022}.}

\par Dispersion interactions between particles and surfaces occur at finite temperatures and in the presence of a background plasma. This is not only of vital importance for various biological applications and catalysis, but it may also surprisingly enough be of interest for hidden aspects in fundamental quantum electrodynamics. All interactions between particles take place in the presence of the plasma of the fluctuating electron-positron pairs; constantly created and annihilated. This is particularly true for the interaction between nuclear particles. Strong similarities were found, suggesting a potential role for screened Casimir forces as one important contribution to the nuclear interaction. When non-relativistic plasma~\cite{EPJDNinham2014} is used, the relativistic energy mc$^2$ enters the interaction energy in a very intriguing way: it replaces the temperature. This indicates that there could be some interesting physics hidden in this problem, and we may need to use the relativistic mass from the beginning. To make further progress, we likely need to extend these simple ideas to include a relativistic plasma response function and to include magnetic (spin) susceptibilities. These are problems of the same importance as occurring in physical chemistry~\cite{NinhamYaminsky1997}.
\revisedtext{A fundamental ansatz commonly used, assumes that all electrostatic interactions (usually analyzed in a nonlinear theory) and electrodynamic interactions (usually treated within a linear approximation) can be treated separately. This in general is in violation of the fundamental physical laws~\cite{NinhamYaminsky1997}. For further progress, one needs to carefully ponder the foundations of the theory of these attractive and repulsive intermolecular interactions\,\cite{PhysRevB.89.201407}. }

\begin{acknowledgments}
 This research is part of the project No. 2022/47/P/ST3/01236 co-funded by the National Science Centre and the European Union's Horizon 2020 research and innovation programme under the Marie Sk{\l}odowska-Curie grant agreement No. 945339. We also thank the "ENSEMBLE3 - Centre of Excellence for nanophotonics, advanced materials and novel crystal growth-based technologies" project (GA No. MAB/2020/14) carried out within the International Research Agendas programme of the Foundation for Polish Science co-financed by the European Union under the European Regional Development Fund, the European Union's Horizon 2020 research and innovation programme Teaming for Excellence (GA. No. 857543) for support of this work.
\end{acknowledgments}

\bibliography{main}

\end{document}